\documentclass[final]{article}

\usepackage[a4paper, total={6in, 8in}]{geometry}
\usepackage{amssymb}
\usepackage{amsmath}
\usepackage{natbib}
\usepackage{graphicx}

\usepackage{xurl}
\usepackage{xcolor}
\newcommand{\BE}{\mathbb{E}}

\title{Think before you fit: parameter identifiability, sensitivity and uncertainty in systems biology models}

\author{Simon P. Preston$^1$, Richard D. Wilkinson$^1$,\\Richard H. Clayton$^2$, Mike J. Chappell$^3$, Gary R. Mirams$^1$}

\begin{document}

\maketitle

\noindent
1. School of Mathematical Sciences, University of Nottingham, University Park, Nottingham, NG7 2RD, Nottinghamshire, United Kingdom\\
2. Insigneo Institute and School of Computer Science, University of Sheffield, Sheffield, S1 4DP, South Yorkshire, United Kingdom\\
3. School of Engineering, University of Warwick, Coventry, CV4 7AL, United Kingdom\\

\begin{abstract}
Reliable predictions from systems biology models require knowing whether parameters can be estimated from available data, and with what certainty. 
Identifiability analysis reveals whether parameters are learnable in principle (structural identifiability) and in practice (practical identifiability). 
We introduce the core ideas using linear models, highlighting how experimental design and output sensitivity shape identifiability. 
In nonlinear models, identifiability can vary with parameter values, motivating global and simulation-based approaches. 
We summarise computational methods for assessing identifiability noting that weakly identifiable parameters can undermine predictions beyond the calibration dataset. 
Strategies to improve identifiability include measuring different outputs, refining model structure, and adding prior knowledge. 
Far from a technical afterthought, identifiability determines the limits of inference and prediction. 
Recognising and addressing it is essential for building models that are not only well-fitted to data, but also capable of delivering predictions with robust, quantifiable uncertainty.
\end{abstract}


\section*{Introduction}
\label{sec1}

The identifiability of parameters from data is an important, but often overlooked, prerequisite to experiment design and parameter estimation.
If parameter estimates are to be used to inform decisions, then it is essential that parameters are identifiable.
Consider a model for  an  experiment where parameters $\theta$ determine a probability distribution $\mathcal{P}_\theta$ for the observed data, $y$. 
{\it Identifiability} of the model is the property that the map $\theta \mapsto \mathcal{P}_\theta$ is injective, that is,  different values of $\theta$ lead to different distributions for the observations, making it possible to `identify' $\theta$ from data. 
If there are two different inputs $\theta_1 \neq \theta_2$ for which 
$\mathcal{P}_{\theta_1}=\mathcal{P}_{\theta_2}$ then the data, $y$, cannot distinguish $\theta_1$ from $\theta_2$, and  the model is said to be {\it unidentifiable} for this experiment.

In this paper, we focus on deterministic models of the form $f(t, \theta)$, usually generated from a differential equation, where $\theta$ is a parameter vector taking values in some admissible set $\Theta\subset \mathbb{R}^p$, and $t$ denotes time. We make the common assumption that observations  $y\in \mathbb{R}^n$ are of the form
\begin{equation}
y = f(\theta) + \varepsilon,
\label{eqn:general:additive:error:model}
\end{equation}
where $\varepsilon$ is a random error, and  $f(\theta)$ is a vector with elements $f(t_i, \theta)$, at time points $\{t_i\}$ called the \emph{design}.

\begin{figure}[htb]
\centering
\includegraphics[width=0.75\linewidth]{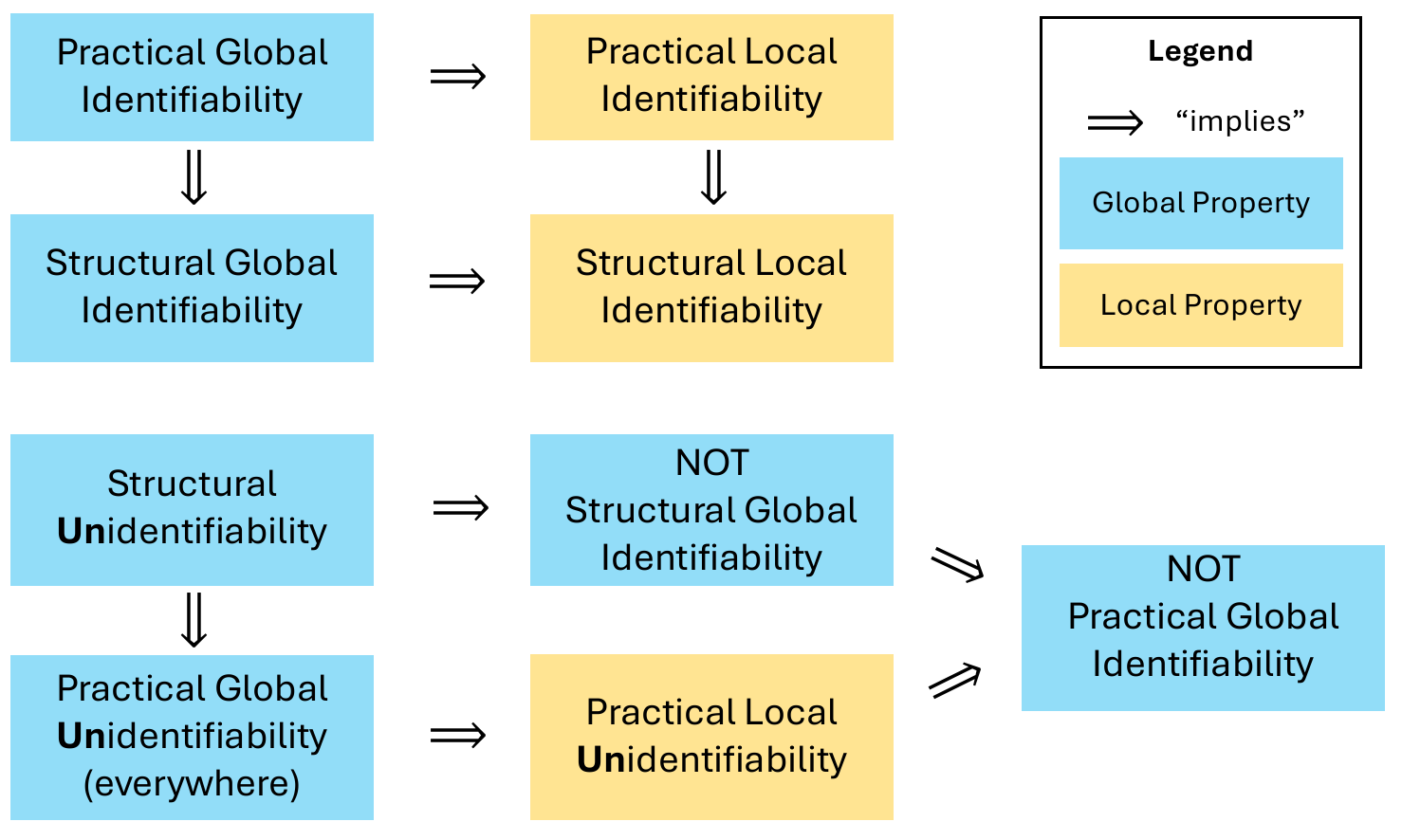}
\caption{A diagram of implication conditions between identifiability concepts discussed.
Note how structural unidentifiability makes practical identifiability impossible, both locally and globally.
Conversely, practical global identifiability, that may be (loosely) tested as described in the `Brute force checks \dots' section, would guarantee the other identifiability conditions.}
\label{fig:implications}
\end{figure}

\section*{Is it mission impossible? Checking structural identifiability}

Structural identifiability addresses the deterministic part of the model, $f$, and asks if we could determine $\theta$ given infinite, noise-free data. 
The answer is no if different parameter sets $\theta, \theta^*$ are
\emph{indistinguishable}, meaning they give rise to identical outputs:
\[
f(t, {\theta}) = f(t, {{\theta}^*}) \mbox{ for all } t.
\]

We say that the model is {\it structurally locally identifiable} at $\theta^*$ if there is a neighbourhood of $\theta^*\!$, $\mathcal N$, for which $\theta$ and $\theta^*$ are distinguishable for 
all $\theta \in\mathcal{N}$.
The model is \emph{structurally globally identifiable} at $\theta^*$ if $\mathcal N = \Theta$, i.e., $f(\theta)\not = f(\theta^*)$ for all $\theta \not = \theta^*$.
If a model is structurally locally identifiable at $\theta^*$ but not structurally globally identifiable at $\theta^*$ (i.e., $\mathcal N$ is strictly smaller than $\Theta$), then $\theta^*$ is distinguishable from other parameters in its neighbourhood, but not from a set of alternatives elsewhere in $\Theta$.
If there is no neighbourhood, $\mathcal N$, for which local identifiability holds then $\theta^*$ is \emph{structurally unidentifiable}, indistinguishable from an 
infinite set of alternative values.
These definitions are in terms of a particular parameter value $\theta^*$, but extend to a \emph{system}, when requiring the respective conditions to hold for all possible $\theta^*\! \in \Theta$.
Relationships between identifiability types are shown in Fig.~\ref{fig:implications}.

How might such unidentifiability arise in a nonlinear system? 
Imagine a model with $f(t, \theta) =
\exp(-\theta_1 t) + \exp(-\theta_2 t)$ and $\theta_1 \neq \theta_2$. Symmetry to swapping $\theta_1$ and $\theta_2$ means that, although the model is locally identifiable, it is not structurally globally identifiable unless $\Theta$ is restricted, e.g.~with $\theta_1 > \theta_2$.
Structural unidentifiability also arises via redundant parameters. 
The model $\theta_1\exp(\theta_2 x + \theta_3)$ is unidentifiable, as it can be written with one fewer parameter, as $\theta_1'\exp(\theta_2x)$, where 
$\theta_1'=\theta_1\exp(\theta_3)$.

In these examples  the unidentifiabilities can be spotted easily and resolved, but in practice we often need to use systematic approaches.
Tools for evaluating the structural identifiability of nonlinear systems include: the Taylor series approach \citep{POHJANPALO197821}, similarity transformation-based approaches \citep{EVANS20021799}, differential algebra techniques \citep{LJUNG1994265,Saccomani2003619,MARGARIA20011}, the observable normal form \citep{Evans201248} and symmetries approaches \citep{Yates2009,Massonis2020469}. Diagnosing unidentifiabilites can be computationally challenging, even for relatively simple models, but recent software offers substantial support. 
Packages include: the Fraunhofer Chalmers Structural Identifiability Analysis tool in Mathematica, 
\citep{anguelova2012minimal,karlsson2012,raue2014comparison}; 
Strike-goldd in Matlab, \citep{Villaverde20181,Massonis20215039,DiazSeoane2023}; 
the COMBOS web app. 
\citep{Meshkat2014}; 
and \texttt{StructuralIdentifiability.jl} in Julia  \citep{dong2023}.
See \citet{heinrich2025structural} for benchmarking. Recently these techniques have been extended to specific forms of spatio-temporal partial-differential equations \citep{renardy2022structural,browning2024structural,salmaniw2025structural,byrne2025algebraic,liu2024parameter} and stochastic differential equations \citep{browning2025exact}.

\section*{Structural identifiability is a minimum requirement}

Structural identifiability refers to the capacity to infer model parameters given  an unlimited supply of perfect, noise-free data, and is hence just a minimum requirement: it is pointless to attempt inference of structurally unidentifiable parameters. 
Instead, one must revisit the model or experimental setup to fix the problem. 
But once structural identifiability is established, \emph{practical identifiability} can be considered.
This takes into account the error term $\varepsilon$ in (\ref{eqn:general:additive:error:model}), asking whether parameters can be inferred with acceptable precision from finite, noisy, potentially sparse, real-world data. 
Consequently, careful `experimental design'---selection of the data to be collected---is essential. 


\section*{Intuition from linear models}

Consider the model $f(\theta)=X\theta$ where data depend \emph{linearly} on $\theta$:  
\begin{equation}
y = X \theta^* + 
\varepsilon \qquad \mbox{with } \varepsilon \sim N(0, \sigma^2 I)
\label{eqn:linear:model}
\end{equation}
for some unknown $\theta^*$.
A natural way to compute an estimate, $\hat{\theta}$, of $\theta^*$ is  to minimise the sum of squares $S(\theta) = \frac{1}{2}\|y - X \theta\|^2$. 
There is a simple closed-form solution, $\hat \theta = (X^\top X)^{-1} X^\top y$, when $(X^\top X)^{-1}$ exists. 

\begin{figure}[htbp]
    \centering
    \includegraphics[width=0.75\linewidth]{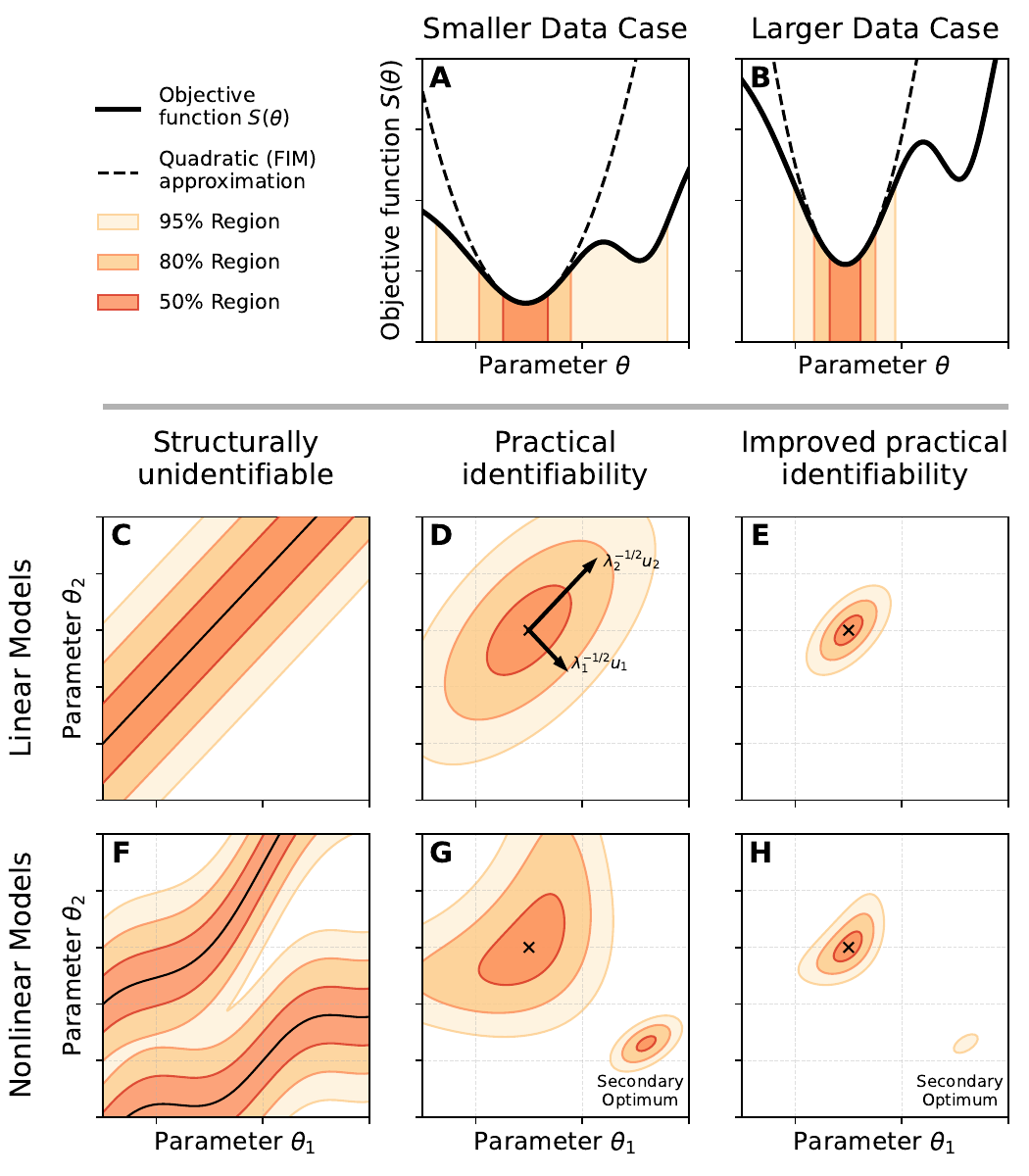}
    \caption{\textbf{A visualisation of identifiability concepts for one (A-B) and two (C-H) parameter dimensions.}
    In all plots, the shading indicates plausible regions for the parameter value, with lighter regions indicating higher confidence, e.g., darkest is 50\% confidence regions, and lightest is 95\% regions. 
    \textbf{Panels A--B: For nonlinear models, the Fisher-information matrix (FIM) provides a better description of parameter identifiability as data size increases.}
    \textbf{A:} The objective function for a nonlinear model, with a secondary minimum. The FIM-based quadratic approximation, which amounts to a linear approximation of the nonlinear model, reflects the 50\% region well, but not the 95\% region which encompasses the secondary minimum. 
    \textbf{B:} With a larger data set, the secondary minimum becomes insignificant; the FIM-based quadratic approximation reflects all the confidence regions with reasonable accuracy.
    \textbf{Panels C--H: cartoon of structural vs.\ practical identifiability in linear and nonlinear cases.}
    \textbf{C\&F:} structurally unidentifiable situation; black lines indicate an infinite set of 
    $\hat{\theta}$ values all of which minimise $S(\theta)$.
    \textbf{D\&G:} the model is structurally globally identifiable, but has poor practical identifiability, as the plausible regions span much of the parameter space. 
    \textbf{E\&H:} structurally identifiable with improved practical identifiability (smaller plausible regions), typical of having more data. 
    \textbf{C--E:} Linear models and identifiability. Practical identifiability properties can be assessed based on the FIM; contours of $S$, and plausible regions, are ellipses centred on $\hat\theta$ (which is shown as a black cross) determined by the eigenvalues/vectors of the FIM. 
    \textbf{F--H:} Nonlinear models and identifiability. 
    As in panels A \& B, nonlinear models mean that 
    $S(\theta)$ is not quadratic in $\theta$. 
    Plausible regions via FIM-based quadratic approximations again miss the secondary optimum. 
    As the data size grows, indicated as we move from panel G to panel H, the plausible regions better resemble those in the linear case. 
    In fact, in the `high data limit' any secondary, less-likely, local optima become insignificant, and the plausible regions computed from the FIM will fully describe the identifiability/uncertainty in the parameter vector $\theta$.
    }
    \label{fig:FIMpluslocalglobal}
\end{figure}

The distribution of $\hat{\theta}$ represents our knowledge/uncertainty about the  true parameter value, $\theta^*$, and should ideally be centred on $\theta^*$ with low variability.
Standard results give that 
\begin{equation}
    \hat\theta \sim N(\theta^*, \mathcal{I}^{-1}),
    \label{eqn:theta:hat:distn:linear:case}
\end{equation} 
where ${\mathcal I} = \sigma^{-2}(X^\top X)\in \mathbb{R}^{p\times p}$. This  distribution is indeed centred on $\theta^*$, and has level sets $\{\theta: (\theta - {\theta}^*)^\top X^\top X(\theta - {\theta}^*) = \operatorname{const}\}$. 
Confidence regions for $\theta^*$ are thus ellipsoids centred on $\hat{\theta}$ with shape determined by $X^\top X$, the Hessian of $S(\theta)$, which describes the curvature/sharpness of $S$; 
see Fig.~\ref{fig:FIMpluslocalglobal}C-E and,  e.g., \citet{bates1988nonlinear} for details.  

$\mathcal{I}$ is the \emph{(Fisher) information matrix} (FIM).
It characterises the `information' $y$ contains about the parameter, $\theta^*$. Consider the eigen-expansion $\mathcal{I} = \sum_i \lambda_i u_i u_i^\top$, where $\lambda_i$ are eigenvalues of $\mathcal{I}$ with corresponding orthonormal eigenvectors, $u_i$.
The eigenvectors, $u_i$, corresponding to larger $\lambda_i$ are the directions in $\Theta$ for which information is large, and vice versa. 
Since $\mathcal{I}^{-1} = \sum_i \lambda_i^{-1} u_i u_i^\top$ determines the variability/uncertainty of $\hat\theta$, directions of high information are directions of low variability for the estimated parameter, $\hat\theta$. 
As the number of data points, $n$, grows, the information $\mathcal I$ typically grows in proportion with $n$, and hence uncertainty about $\hat{\theta}$ shrinks --- see 
 Figure \ref{fig:FIMpluslocalglobal}D and E.
 
The matrix $X$ is often called the \emph{design matrix}, as it defines what data are collected. {\it Experimental design} is the process of  choosing a `good' $X$. 
Since $\mathcal{I}$ is proportional to $X^\top X$, we can see that the design $X$ determines the information in $y$ about $\theta$. 
$X$ can also be interpreted as a `sensitivity matrix' since $\frac{{\rm d }\BE y}{ {\rm d} \theta} = X$, so it determines how sensitive the data $y$ are to changes in $\theta$. 
Directions for which $\BE(y)$ is insensitive to changes in $\theta$ are those in which the data $y$ possess little information about $\theta$. 

If the information matrix $\mathcal I$ has a zero eigenvalue (i.e.~is `rank deficient', equivalently  $\text{det} \, \mathcal I = 0$), then the linear model is unidentifiable---and vice versa (Fig.~\ref{fig:FIMpluslocalglobal}C). 
In this case $(X^\top X)^{-1}$ does not exist, and there is not a unique minimiser of  $S(\theta)$. 
This always occurs when we have fewer data points than parameters (the {\it underdetermined case}), but can also result from poor experimental design $X$.

\section*{From linear to nonlinear}

If $f(\theta)$ is nonlinear in $\theta$ we can still minimize the sum of squares $S(\theta) = \frac{1}{2}\|y - f(\theta) \|^2$ to obtain the least-squares estimator $\hat{\theta} = \text{argmin}_\theta S(\theta)$.
But now $\hat{\theta}$ does not have a closed-form solution; we must resort to numerical optimisation. 
Concepts can be connected to the linear case by considering a Taylor expansion
\begin{equation}
y = f(\theta^*) + V(\theta^*) (\theta - \theta^*) + \ldots + \varepsilon,
\label{eqn:linearised:model}
\end{equation}
where  $V(\theta^*) = \frac{{\rm d}f}{{\rm d}\theta}\big|_{\theta^*}$ is the $n$-by-$p$ sensitivity matrix evaluated at $\theta^*$, and `$\ldots$' indicates terms involving higher powers of $(\theta - \theta^*)$ (that we'll neglect).
If $\theta^*$ is the `true' value of the parameter, then
$\frac{{\rm d }\BE y}{ {\rm d} \theta} \big|_{\theta^*}  = V(\theta^*)$. 
Thus  $V$  plays the same role here as the design matrix $X$ in the linear case, but now varies with $\theta^*$.

Under commonly holding conditions (loosely speaking, that $f$ is smooth, $\Theta$  compact, and $\theta^*$ is not on the boundary of $\Theta$---see e.g.~\citep{SeberWild1989}) then $\hat{\theta} \rightarrow \theta^*$ with the distribution
\begin{equation}
\hat{\theta} \sim N\left(\theta^*, \mathcal{I}(\theta^*)^{-1}\right)
\label{eqn:distn:of:theta:hat:nonlinear:case}
\end{equation}
as $n \rightarrow \infty$, where $\mathcal I(\theta^*)=\sigma^{-2}V(\theta^*)^\top V(\theta^*)$. 
When $f$ is nonlinear and $n$ is finite, this distribution is only an approximation, but is often adequate.
The information/uncertainty about $\theta^*$ can be visualised using  \eqref{eqn:distn:of:theta:hat:nonlinear:case} as an ellipsoid determined by $\mathcal I(\theta^*)$, which plays the same role as for the linear case: 
its eigenvalues and eigenvectors provide directions in $\Theta$ for which the data are informative about $\theta^*$. 
As $n$ grows, the distribution of $\hat{\theta}$ gets increasingly concentrated, such that only the region around $\hat{\theta}$ is relevant; 
hence ellipsoidal levels sets defined by $\mathcal I (\hat{\theta})$ get better and better at describing the uncertainty about $\theta^*$. Fig.~\ref{fig:FIMpluslocalglobal} illustrates this for 1D and 2D cases.

\begin{figure}[tbhp]
    \begin{center}
    \includegraphics[width=0.75\linewidth]{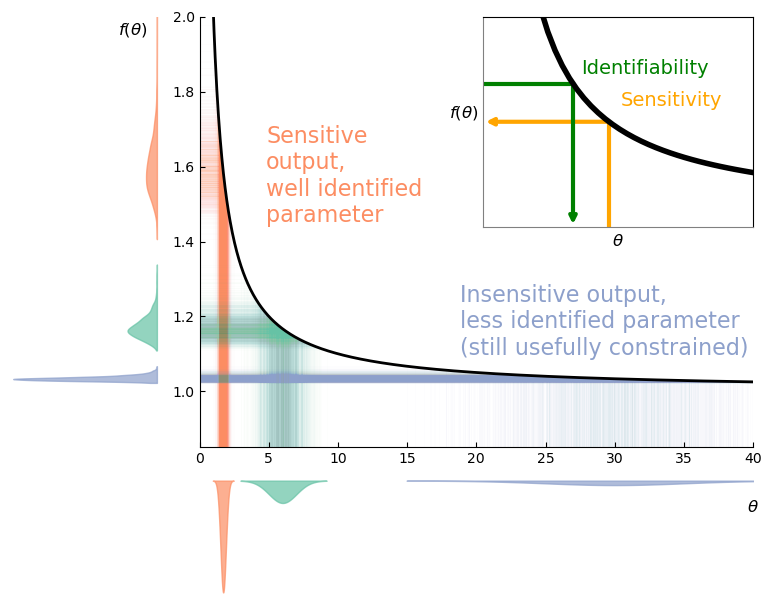}
    \caption{\textbf{Sensitivity depends on the parameter values in nonlinear models and predictive uncertainty is not the same as parameter uncertainty}. 
    A graph of the function $f(\theta)=1 + \frac{1}{\theta}$. We highlight how parameter uncertainty can have a nonlinear relationship with uncertainty in $f(\theta)$ (predictive uncertainty).
    Identifiability analysis is the study of what we can learn about $\theta$ from observing $f(\theta)$.
    Sensitivity analysis is the study of how $f(\theta)$ varies depending on $\theta$.
    Scenarios where a model has poor parameter identifiability but constrained output uncertainty may be perfectly acceptable, but only if we can show that the outputs under intended experimental designs (for particular contexts of use) remain insensitive and do not become subject to high predictive uncertainty.}
    \label{fig:nonlinear_func}
    \end{center}
\end{figure}

To summarise, $\mathcal{I}(\theta)$ determines the  information about $\theta^*$ and its local identifiability. A model is locally structurally identifiable at $\theta$ if and only if $\text{det} \,\mathcal I(\theta)\not = 0$. 
A full-rank FIM implies that all parameters distinctly affect $y$, and can hence be learned from $y$.
In the nonlinear case, 
$\det{\mathcal{I}(\theta)}$ depends on $\theta$, but in the linear case $\mathcal{I}$ is constant and hence $\text{det} \, \mathcal I \neq 0$ implies both local and global identifiability.

In the nonlinear case, just as the linear case, identifiability and sensitivity are two sides of the same coin;
$V(\theta) =  \frac{{\rm d \BE y}}{{\rm d}\theta}$ is a  sensitivity matrix determining how $y$ changes with $\theta$. If there are changes in $\theta$ to which $y$ is insensitive, then it will  be difficult to identify $\theta$ from $y$.

\section*{Practical identifiability}

\emph{Practical} identifiability broadly means that the uncertainty in the inferred parameters is acceptably low, which can again be determined from the FIM $\mathcal{I}(\theta)$.
Supposing we establish that $\mathcal I(\theta)$ has no zero eigenvalues (and thus the model is locally structurally identifiable), the next question is whether its smallest eigenvalues are large enough that, via
\eqref{eqn:distn:of:theta:hat:nonlinear:case},
the uncertainty in  $\hat{\theta}$ is acceptably small.

If only some linear combination, say $a^\top \theta^*$, of $\theta^*$ is of interest it can be practically identifiable when $\theta^*$ is not. 
From \eqref{eqn:distn:of:theta:hat:nonlinear:case},
\begin{equation}
a^\top\hat{\theta} \sim N\left(a^\top \theta^*, a^\top \mathcal{I}(\theta^*)^{-1} a\right),
\label{eqn:linear:combo:normal}
\end{equation}
so if $a$ aligns, say, with an eigenvector corresponding to larger eigenvalues of $\mathcal I(\theta^*)$, then the variance of $a^\top\hat{\theta}$ can be small even if  the variance of  $\hat{\theta}$ is large. 
When $a$ is a vector with the $i$th element equal to $1$, and zeros for the rest, then \eqref{eqn:linear:combo:normal} defines the distribution for the estimate of a single parameter, $\hat{\theta}_i$.

Closely related is \emph{sloppiness}, a term introduced to describe models with eigenvalues of $\mathcal{I}$ approximately linearly spaced on the log scale, something widely observed in nonlinear systems biology models \citep{gutenkunst2007universally}.
Sloppy models tend to have a few (combinations of) parameters, termed `{stiff}', which determine  the model behaviour (and are thus easily estimated from data), and other parameter (combinations) terms `sloppy' which are practically unidentifiable. 

\section*{Challenges of nonlinear models}

Consider the model in Fig. \ref{fig:nonlinear_func}
\begin{equation}
f(\theta) = 1 + \frac{1}{\theta}.
\end{equation}
When $\theta$ is small it is easily identifiable, but becomes unidentifiable as it grows. 
If our interest is in calibrated prediction, i.e., the uncertainty in $f(\hat{\theta})$, then inability to identify $\theta^*$ may not matter if predictive uncertainty remains small. 
This illustrates the importance of defining the quantity of interest early. 
Care is needed when the fitted model is to be used predictively outside the context of the training data: parameters that are unidentifiable and unimportant in the training context can become critical at prediction time \citep{whittaker2020calibration}.
 
\subsection*{Can we average?}

In complex nonlinear models, we may not have access to gradient information $\nabla f$, and $f(\theta)$ may be computationally expensive to evaluate, making computation of the FIM $\mathcal{I}(\theta)$ challenging. 
Global sensitivity analysis (GSA) methods summarise the overall importance of different inputs across $\Theta$, i.e., rather than providing an at-a-point definition of sensitivity, they seek a global or average quantification of the importance of each parameter. 
\textit{Sobol indices} are widely used for GSA to summarise the value of learning each different parameter in terms of reducing output uncertainty.
The {\it prior} uncertainty about $\theta$, described by a probability distribution $\pi(\theta)$, is  `{pushed forwards}' 
through the model to give the output uncertainty $\text{Var} f(\theta)$.
The \emph{first-order indices}, $S_i$,  give the expected variance reduction in the output if we knew the true value of $\theta_i$. The \emph{total-order indices}, $S_{T_i}$, give how much uncertainty would remain if we knew all inputs except $\theta_i$, written $\theta_{\sim i}$:
\[
S_i = \frac{\text{Var}\big(\BE(y|\theta_i)\big)}{\text{Var}(\,y\,)}, \qquad S_{T_i} = 1-\frac{\text{Var}\big(\BE(y|\theta_{\sim i})\big)}{\text{Var}(\,y\,)}.
\]
Non-zero Sobol indices suggest potential identifiability, but do not guarantee it. 
Conversely, if $S_i, S_{T_i}\approx 0$, reliable estimation of $\theta_i$ is unlikely.

The advantage of Sobol indices \citep{sobol_global_2001, saltelli_variance_2010} and alternative approaches (including Morris screening \citep{morris_factorial_1991}, Shapley values \citep{owen2014:shapley}, FAST \citep{homma1996importance}), is that they can typically be computed even for complex models where the FIM is unavailable. 

\subsection*{Emulators}

When $f(\theta)$ is computationally expensive, a common strategy is to use an \emph{emulator}: a surrogate model trained on a limited  budget of model evaluations that mimics $f(\theta)$. 
Emulators are computationally inexpensive and can be used in analyses, such as GSA, that would otherwise be impossible. 
Gaussian processes are a very popular choice of emulator \citep{gramacy} as they estimate the approximation uncertainty, but methods such as linear regression, 
polynomial chaos expansions \citep{gerritsma_time-dependent_2010}, 
and neural networks 
are also widely used. 

\subsection*{Other global methods}

\emph{Profile (log)likelihood} functions can be used to assess both  structural and practical identifiability \citep{kreutz2013profile, wieland2021structural} of an individual parameter $\theta_i$. 
For models with additive Gaussian errors, the log-likelihood $\ell(\theta)$  is proportional to $-S(\theta)$.
The profile log-likelihood fixes one parameter value, say $\theta_i$, and `profiles out' the others by maximising (locally or globally \citep{venzon1988method}) $\ell(\theta)$ with respect to them:
\begin{displaymath}
p_i(\theta_i) = \max_{\theta\, \vert \, \theta_{i} \,\text{fixed}} \ell(\theta).
\end{displaymath}
Structural unidentifiability appears in $p_i(\theta_i)$ as a completely flat region around its maximum, whereas poor practical identifiability appears as a relatively flat region around the maxima.

{\it Minimally disruptive curves} \citep{raman2017} 
are a generalisation of profile likelihood computed by finding paths through parameter space that preserve model behaviour, illustrating the unidentifiabilities between parameters, and suggesting model simplifications to improve identifiability.
Both methods rely on efficient optimization, which may not be possible for complex models.

\section*{Brute force checks with synthetic data}

An important check is to generate \emph{synthetic data} using some $\theta$, before then 
`forgetting' $\theta$ and attempting to re-infer it from the data.
This tests both practical identifiability and robustness of the optimisation/inference algorithm whose success often depends on multiple factors such as algorithm hyperparameters and parameter transformations \citep{owen2025ionbench, whittaker2020calibration}.
If the goal is one-off calibration, local identifiability of the true parameter set $\theta^*$ is sufficient. 
But when developing a pipeline to fit many different future datasets, we will need to evaluate global identifiability using many different $\theta^*$ sampled widely from $\Theta$.
Repeated success, with noise and time sampling appropriate to the experimental setting, can give confidence that we have practical global identifiability.

An alternative to optimising an objective function to obtain $\hat\theta$, is \emph{sampling} approaches, to explore the landscape of plausible $\theta^*$, as in Fig.~\ref{fig:FIMpluslocalglobal}G--I. 
This is common in Bayesian approaches, which estimate a posterior distribution for parameters, often using  Markov chain Monte Carlo methods \citep{siekmann2012mcmc, hines2014determination}. 
However these can require $\mathcal{O}(10,000)$ or more evaluations of $f(\theta)$---although emulators can reduce computational cost. 
Methods such as history matching \citep{Vernon, Craig} and approximate Bayesian computation \citep{csillery2010, wilkinson2013} are also useful tools for checking practical identifiability, and can incorporate an assessment of model discrepancy without requiring a full probabilistic model to be specified. 
See \citep{wilkinson2023introduction} for a recent review of calibration methods for complex models.


\begin{figure}
    \includegraphics[width=\linewidth]{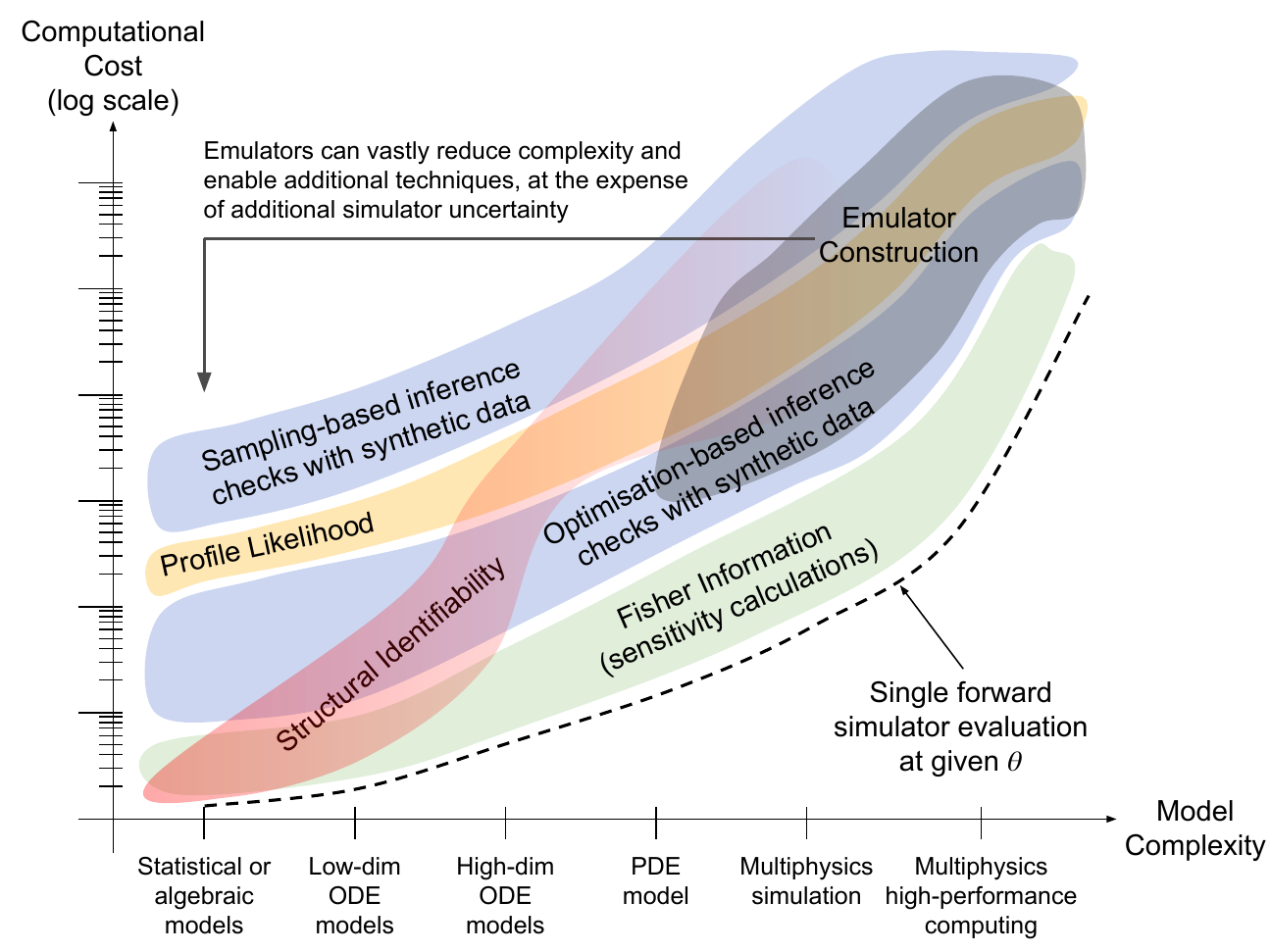}
    \caption{A schematic showing some techniques for assessing identifiability and how computationally expensive they are for computational models of different types, ranked in a typical order of computational complexity.}
    \label{fig:workflows}
\end{figure}

\section*{Improving parameter identifiability}

Identifiability  can  be improved by simplifying the model, enhancing the experiment, or both \citep{apgar2010sloppy, jeong2018experimental}.
Enhancing the experiment might involve  observing more model `states' ---  the notion of {\it minimal output sets} for ensuring structural identifiability can help when there is a choice of observables \citep{joubert2018determining}, choosing suitable time points for observations, and  optimising experimental conditions --- for example, control of a forcing function to perturb system dynamics to regions of higher parameter sensitivity. 
Repeated measurements improve identifiability by reducing the variance in parameter estimates.
The field of {\it optimal experimental design} is extensive, 
and classical approaches are based on determining designs that optimise scalar functions (such as the determinant or trace) of the sensitivity or Fisher information matrices \citep{goodwin1977dynamic}.
{\it Generalised sensitivity functions} \citep{thomaseth1999generalized, fink2009markov}, or {\it information sensitivity functions} \citep{pant2018} can be helpful to show which parts of an experiment provide most information on each parameter.

The simplest way to simplify the model is to fix unidentifiable parameters to reference values: domain experts usually have intuition for relative parameter sizes, and GSA can be used to choose which parameters to fix.
Whilst not improving identifiability of parameters as such, physically informed priors and/or regularisation are used commonly to overcome flat objective functions and improve algorithm performance \citep{Munch2024,semochkina2025incorporating}.

Asymptotic analysis can reduce the model by omitting processes (and associated parameters) that are too slow or fast to be evident in the data \citep{biktasheva_asymptotic_2006}.
{\it Manifold boundary approximation} \citep{transtrum2014model} is an approach to model reduction, successfully applied in systems biology modelling, \citep{jeong2018experimental,whittaker2022ion}, that moves parameters in a direction which minimises change in the model output until eventually reaching a limit where they become very small or large without influencing the output further. 
A model reduction step involves removal or lumping to reduce the number of parameters, making a simpler model that is restricted to one boundary of the more complex model's behaviour. 
After any model reduction, care should be taken to check fit and prediction quality and any change in predictive uncertainty.

\section*{Computational considerations}

Fig.~\ref{fig:workflows} shows how the computational cost of identifiability techniques can increase with the complexity of the computational model.
Fisher-information-based uncertainty around a proposed parameter set is often the simplest measure of (local) identifiability. 
The required sensitivities $\partial f/\partial \theta$ can be computed in a variety of ways including finite difference approximations (although care needs to be taken to ensure appropriate step size \citep{creswell2024understanding}), developing an augmented set of  ODEs and simultaneously solving with the model \citep{cobelli1980parameter}, and using automatic differentiation software which can compute derivatives for any numerical quantity
\citep{margossian2019review} --- which can work seamlessly in Julia due to its multiple-dispatch and flexible types \citep{sapienza2024differentiable}. 
For more intensive methods, such as synthetic data checks and profile-likelihood, emulators may be needed for more computationally expensive models \citep{colebank2025assessing}.


\section*{Conclusions}

Identifiability checks are a key part of any workflow when building robust models with well-characterised uncertainty capable of reliable quantitative predictions.
We have outlined techniques ranging from local analysis provided by the FIM, to global methods including profile likelihood and synthetic data checks.
If a model has poorly identifiable parameters, we can either enhance the experimental design to collect more informative data, or simplify the model so that the parameters are better informed by the data.
Systems biology models inevitably increase in complexity as we add features to achieve better data-fits \citep{white2016limitations}, but often with diminishing returns and the effect of making some parameters poorly identifiable.


Beyond the scope of this review is \emph{model discrepancy/misspecification}, \emph{i.e.}, where  
even the `best' $\theta^*$ cannot perfectly (i.e.\ within statistical error) reproduce the data. 
This means parameters will not, in practice, be calibrated with the fidelity that identifiability checks suggest, as the calibrated parameters are ultimately those of the model, not necessarily true physical values. 
Parameter estimates will be biased as they must compromise to get the best fit for a misspecified model  \citep{tuo2015,lei2020considering,shuttleworth2024empirical}. 
We can try to develop a statistical model of the model discrepancy/error \citep{Kennedy}, but this often introduces unindentifiabilities into the model \citep{brynjarsdottir2014learning, plumlee2017}. 
These observations reinforce that predictive checks against unseen validation data are vital for assessing a calibrated model's real predictive power.

\section*{Acknowledgments}

This work was supported by the Wellcome Trust (grant no. 212203/Z/18/Z); and the EPSRC (grant nos. EP/Z531297/1, EP/X012603/1, EP/Z000580/1, EP/W000091/2). GRM acknowledges Wellcome's support via a Senior Research Fellowship. 
The authors would like to thank the Isaac Newton Institute for Mathematical Sciences, Cambridge, for support and hospitality during the programme ``Representing, calibrating \& leveraging prediction uncertainty from statistics to machine learning'', where work on this paper was undertaken. 



\end{document}